\documentclass[12pt, thmsa]{article}
\usepackage{pgfpages}
\usepackage{soul}
\usepackage[polutonikogreek, english]{babel}
\usepackage{fullpage}
\usepackage{setspace}
\usepackage{graphicx}
\usepackage{latexsym}
\usepackage{amsmath}
\usepackage{amstext}
\usepackage{euscript}
\usepackage{epsfig}
\usepackage{color}
\usepackage{ulem}
\usepackage{soul}
\usepackage {setspace}

\usepackage{enumerate}
\usepackage{refcount}
\usepackage{mdwlist}
\usepackage{fixltx2e}

\usepackage[T1]{fontenc}
\usepackage{amssymb}
\usepackage{lmodern}
\usepackage{ragged2e}

\begin{document}

\title{The Agnostic Structure of Data Science Methods}

\author{Domenico Napoletani\footnote{University Honors Program and Institute for Quantum Studies, Chapman University, Orange (CA)}, Marco Panza\footnote{CNRS, IHPST (CNRS and Univ. of Paris 1, Panth\'{e}on-Sorbonne) \& Chapman University, Orange (CA)
}, Daniele Struppa\footnote{The Donald Bren Presidential Chair in Mathematics, Chapman University, Orange (CA)}}

\date{}

\maketitle

\begin{abstract}
\noindent
In this paper we argue that data science is a coherent and novel approach to empirical problems that, in its most general form, does not build understanding about phenomena. Within the new type of mathematization at work in data science, mathematical methods are not selected because of any relevance for a problem at hand; mathematical methods are applied to a specific problem only by `forcing', i.e. on the basis of their ability to reorganize the data for further analysis and the intrinsic richness of their mathematical structure. In particular, we argue that deep learning neural networks are best understood within the context of forcing optimization methods. We finally explore the broader question of the appropriateness of data science methods in solving problems. We argue that this question should not be interpreted as a search for a correspondence between phenomena and specific solutions found by data science methods; rather, it is the internal structure of data science methods that is open to precise forms of understanding.
\end{abstract}

\section{Introduction}
In this paper we want to discuss the changing role of mathematics in science, as a way to discuss some methodological trends at work in big data science.  Classically, any application of mathematical techniques requires a previous understanding of the phenomena and of the mutual relations among the relevant data. Modern data analysis, on the other hand, does not require that. It rather appeals to mathematics to re-organize data in order to reveal  possible patterns uniquely attached to the specific questions we may ask
about the phenomena of interest. These patterns may or may not provide further understanding \textit{per se}, but nevertheless provide an answer to these  questions.

It is because of this diminished emphasis on understanding that we suggested in \cite{nps1} to denote such methods under the label of `agnostic science', and we speak of `blind methods' to denote individual instances of agnostic science. These methods usually  rely only on large and diverse collections of data to solve questions about a phenomenon. As we will see in Section 3, a reliance on large amounts of data is, however, not sufficient in itself to make a method in data science blind. 

The lack (or only partial presence) of understanding of phenomena in agnostic science makes the solution to any  specific problem  dependent on patterns identified automatically through mathematical methods. At the same time, this approach calls for a different kind of understanding of what makes mathematical methods and tools well adapted to these tasks.

A current explanation of the power of data science is that it succeeds only because of the relevance and size of the data themselves. Accordingly, the use  of mathematics in data science reduces itself to disconnected methods devoid of a common structure. This view would amount to a new Wignerian paradox of ``unreasonable effectiveness'', where such effectiveness is assigned to data and not to mathematics. This affirmation of the exclusive primacy of data would be a sort of revenge of facts against their mathematization.

We reject such an answer as both immaterial and unsupported. In our rejection we do not argue that any exploration of data is doomed to failure without some previous understanding of the phenomenon (in other words, we do not oppose data science's methods by defending the classical approach). Rather we observe the effectiveness of agnostic science in the absence of previous understanding, and wonder about what makes this possible.

While we do not advance here any comprehensive reason for the effectiveness of data science, we observe that no account is possible unless we engage in a technical inquiry on how these algorithms work, and suggest a largely schematic account of their \textit{modus operandi}.  This account relies on the results of our previous works \cite{nps1}, \cite{nps4},  \cite{nps5}, which we reorganize in a comprehensive way. Furthermore, we identify a possible direction for future researches, and a promising perspective from which the question can be tackled.

In \cite{nps1} we discussed the lack of understanding proper to the current big data methods. In so doing, we did not borrow any general and univocal notion of
understanding for an empirical (physical, biological, biomedical, social, etc.) phenomenon. We simply observed that big data methods typically apply not to the study of a certain empirical phenomenon, but rather to a pair composed by a phenomenon and a question about it. 

Such methods are used when it is impossible to identify
a small number of independent variables whose measurement suffices to describe the phenomenon and to answer the question at hand. 

We have argued that, when this happens, no appropriate
understanding of the phenomenon is available, regardless of how one could conceive the notion of understanding. This highlights the 
need to understand why mathematics is successful in agnostic science successful (and when), despite the blindness of its methods.

Let us now briefly describe the structure of the paper. 

In Section \ref{sect.MP}, we discuss what we consider to be the basic trend of agnostic science: the ``Microarray Paradigm'', as we called it in \cite{nps1}. 
This name was chosen to reflect the fact that this trend  became first manifest in biology and biomedical sciences, though it is now pervasive in all data science. It is  characterized by the handling of huge amounts of data,
whose specific provenance is often unknown, and whose modes of selection are often disconnected from any previous identification of a relevant structure in the
phenomenon under observation. This feature is intended as the most notable virtue of the paradigm, since it allows investigating the relevant phenomena, and specifically
the data that have been gathered about them, without any previous hypothesis on possible correlations or causal relationships among the variables that are measured.

As already noted, there is an important distinction to be done between the use of powerful and often uncontrollable algorithms on huge amounts of data and agnostic science properly said.  This distinction will be investigated
in Section \ref{sect.ASINS} with the help of a negative example, the Page Rank algorithm used by Google to weight web pages. The basic point, here, is that the lack of local control of the algorithm in use is not the same as lack of understanding of the relevant phenomenon. The former is proper to any algorithm working on huge amounts of data, like  Page Rank; the latter  is instead, by definition,  the characteristic feature
of agnostic science.

In Section \ref{sect.FOR}, we will go further in our analysis of agnostic science, by investigating the relations between optimization and ``forcing'', a term we first introduced in this context in \cite{nps1}.  More specifically, by forcing we referred in \cite{nps5} to the following methodological practice:
\begin{quote}
{\it The use of specific mathematical techniques on the available data that is not motivated by the understanding of the relevant phenomena, but only by the ability of such techniques to structure the data to be amenable to further analysis.} 
\end{quote}
\noindent
For example, we could impose continuity and smoothness on the data, even in the case of variables that can only take discrete value, just for the purpose of being able to use derivatives and differential equations to analyze them. In cases such as this, we say that we are forcing mathematics over the available  data.

In our terminology, optimization itself can be seen as a form of forcing when we carefully consider the ways it is used within agnostic science.

By better describing this \textit{modus operandi}, in relation to deep learning techniques, we will also make clear that the reasons for the effectiveness of optimization cannot be regarded as
the rationale for the success of agnostic science. In short, this is because optimization is itself a form of forcing and does not ensure that  the optimal solutions found by such methods correspond to anything of significance in the evolution or state of the phenomenon. Having made this clear, in Section \ref{sect.Con}, we shall outline a tentative answer to the question of the success and the appropriateness of agnostic science, by indicating a possible direction for further reflection. 

Our suggestion is based on the observation that blind methods can be regarded as complying with a simple and general prescription for algorithms' development, what we called ``Brandt's Principle'' in \cite{nps5}. We then discuss the way in which data science (\cite{Hastie2016}, Chapter 16) relies on  entire families of methods (``ensemble methods''), and we interpret it in light of the Microarray Paradigm, ultimately showing how this forces some implicit analytic constraints on blind methods. We finally propose that  these analytic constraints, when taken together with Brandt's principle, exert strong restrictions on the relevant data sets that agnostic science can deal with.

\section{The Microarray Paradigm and Supervised Learning \label{sect.MP}}

We now describe in some detail a biological problem and a corresponding experimental technique. This provides a paradigmatic example of a blind method, and illustrates the typical way agnostic science works.

One of the great advances of biology and medical practice has been the understanding of the relevance of genes in the development of several diseases such as cancer. The impact of genetic information, as encompassed by DNA sequences, is primarily mediated in an organism by its expression through corresponding messenger RNA (mRNA) molecules. 
We know that the specific behavior of a cell largely depends on the activity, concentration, and state of proteins in the cell, and the distribution of proteins is, in turn, influenced by the changes in levels of mRNA. This opens the possibility of understanding diseases and their genetic basis through the analysis of mRNA molecules.

The mechanism that leads from a certain distribution of mRNA molecules to the manifestation of a certain disease is, however, often not understood, as well as not understood are the specific mRNA molecules that are relevant for particular diseases.
Biologists developed a technique, called `DNA microarray', that can bypass, to some extent, this lack of understanding, and allows the identification of patterns, within mRNA distributions, that may be markers of the presence of some diseases.

We first describe very briefly the experimental structure of the DNA microarray, and then the way it can be used in diagnostics (we refer to \cite{nps1} for a list of references on this technique, or to \cite{DNAmicroarray} for a broad introduction).

A DNA microarray is essentially a matrix of microscopic sites where several thousands of different short pieces of a single strand of DNA  are attached. Messenger RNA  (mRNA) molecules are extracted from some specific tissues of different patients, then amplified and marked with a fluorescent substance and finally dropped on each site of the microarray. 

This makes each site take a less or more intense florescence
according to the amount of mRNA that binds with the strands of DNA previously placed in it. The intensity and distribution of the fluorescence give a way to evaluate the degree of complementarity
of the DNA and the mRNA strands.

This provides a  correspondence between the information displayed by a  DNA microarray and the behavior of a cell. This correspondence, however, is by no means exact
or univocal, since the function of many proteins in the cell is not known, and several strands of DNA  are complementary to the mRNA strands of all protein types. Nevertheless, thousands of strands of
DNA are checked on a single microarray, so that one might expect this method to offer a fairly accurate  description of the state of the cells, even if it does not offer any 
sort of understanding of what is happening in the
relevant tissues. The microarray is, indeed, particularly valuable for a huge number of variables, whose relation to each other and to the state of the cell we ignore.

This does not forbid, for example, the use of microarrays for the diagnosis of many illnesses,  since by measuring the activity of proteins one may be able to distinguish patients affected by a certain pathology from those patients that do not, even without knowing the reason for the differences. 

From a mathematical point of view, this process can be described as follows: let us  label by `$X_i$' the vector of expression levels of mRNA strands associated to a patient $i$: the hope is to be able to find a function $F$ such that $F(X_i)=0$ if the patient does not have a particular disease, and $F(X_i)=1$ if the patient does have the disease. The question of how to find such a function $F$ is at the core of agnostic science, and we will come back to it momentarily. 

This short description of DNA microarrays should already be enough to
justify why this technology can be taken as a paradigmatic example of the way agnostic science works. In short, this point of view can be summarized as follows: 
\begin{quote}
{\it If enough and sufficiently diverse data are collected regarding a certain phenomenon, we can answer all relevant questions about it.}
\end{quote}\noindent This slogan, to which we refer as the Microarray Paradigm, pertains not only to this specific example, but applies as well to agnostic science as a whole\footnote{We introduced the term `Microarray Paradigm' in \cite{nps1} to refer to a specific attitude towards the solution of data science problems, and not only to denote the technique underlying DNA microarrays.  The microarray paradigm, as an attitude to the solution of problems, should not be confused with agnostic science, which is a general scientific practice implementing  this attitude. It should also not be confused with any particular blind method---namely a specific way to implement the microarray paradigm.}. The question we want to tackle here is what makes this paradigm successful (at least in a relevant number of cases).

To better grasp the point, let us sketch the general scheme that agnostic science applies under this paradigm: data are processed through an appropriate algorithm that works on the available data independently both of their specific nature, and of any knowledge concerning the relations possibly connecting the relevant variables. The process is subject to normalization constraints  imposed by the data, rather than by  the (unknown) structure of the phenomenon, and this treatment produces an output which is taken as an  answer to a specific question about this phenomenon.

This approach makes it impossible to generalize the results or even to deal with a change of scale: different questions require different algorithms, whose structure is general and applied uniformly across different problems. Moreover, the specific mathematical way in which a question is formulated depends on the structure of the algorithm which is used, and not the other way around. 

To better see how blinds methods work, we will provide a quick overview of  Supervised Machine Learning  (we shall came back later to this description in more
detail). While this is just an example, it will show the strong similarity between blind methods  and interpolation and approximation theory.

The starting point is a training set $\left(X,Y\right)$, constituted by $M$ pairs $\left(X_{i},Y_{i}\right)$ ($i = 1,2, \ldots , M$), where each $X_{i}$  is typically an array $\left(X^{[j]}_{i}\right)$
($j = 1,2, \ldots , N$) of given values of $N$ variables.
For example, in the DNA microarray example,  each array $X_i$ is the expression of mRNA fragments detected by the microarray, while the corresponding $Y_{i}$ indicates the presence ($Y_i=1$) or absence ($Y_i=0$) of a given disease.

By looking at this training set with the help of an appropriate algorithm, the data scientist looks for a function $F$ such that $F(X_{i}) =Y_{i}$ or  $F(X_{i}) \approx Y_{i}$ ($i = 1,2, \ldots , M$). This function is usually called a `classifier' if the output is categorical, a `model' or a `learned function' for continuous outputs. In the following we will refer in both cases to $F$ as a `fitting function', to stress the connection of supervised machine learning with approximation theory.
 
In general, one looks for a function $F$ that satisfies these conditions for most indices $i$; moreover it is helpful if the function belongs to a functional space $\mathcal{F}$ selected because of its ability to approximate general regular functions in a computationally efficient way. For example, if the regular functions of interest are analytical functions over an interval, $\mathcal{F}$ can be taken to be the space of polynomial functions defined on the same interval.

Let $\mathcal{A}$ be the space of parameters that define a function in $\mathcal{F}$, and denote its elements as $F_a(X)$', where $a$ is an array of parameters in  $\mathcal{A}$. The standard way to find the most suitable $F_a\in\mathcal{F}$ for a supervised learning problem is akin to functional approximation that can be described as follows. One defines a ``fitness function'' $E(a)$ by comparing the output $F_a(X_i)$ to the corresponding value $Y_i$ at each $X_i$, and by setting $$E(a)=\sum_i (Y_i-F_a(X_i))^2.$$
The most suitable $F_{\bar a}$ is identified, then, by seeking a value $\bar a$ that minimizes $E(a)$. The function $F_{\bar a}$ selected in this way is, then, tested on a testing set $\left(X_{i},Y_{i}\right)$, ($i = M+1, M+2, \ldots M+M'$), and, if it is found to be accurate on this set as well, is finally used, by analytical continuation, to forecast $\tilde Y$ when a new instance of argument $\tilde X$ is taken into account.

Though brief, this description shows that supervised machine learning consists in analytically continuing a function found by constraining its values on a discrete subset of points defined by the training set.
Moreover, supervised learning gives a mathematical form to the basic classification problem: given a finite set of classes of objects, and a new object that is not labelled, find the class of objects to which it belongs. What is relevant, however, is that each of the numbers $M,$ $N$, and $M'$ may be huge, and we have no idea of how the values $Y_{i}$ depend on the values $X_{i}$, or how the values in each array  $X_{i}$  are related to each other. In particular, we do not know whether the variables
taking these values are reducible (that is, depend on each other in some way) or whether it is possible to apply suitable changes of scale or normalization on the variables.

Supervised learning is therefore emblematic of agnostic science since we have no way to identify a possible interpolating function $F_a$, except the use of appropriate algorithms. Our lack of understanding of the phenomenon, in particular ensures that there is no effective criterion to guide the choice of the vector of parameters $a$, which are instead taken first to be arbitrary values, and eventually corrected by successive reiterations of the algorithm, until some sort of stability is achieved.

Note that not all data science algorithms fall directly under the domain of supervised learning. For example, in unsupervised learning, the goal is not to match an input $X$ to an output $Y$, but rather to find patterns directly in 
the set of inputs $\{X_1,\ldots,X_M\}$.  The most recent consensus is that unsupervised learning is most efficiently performed when conceived as a particular type of supervised learning (we shall come back later to this point). Another important modality of machine learning is reinforcement learning, a sort of semi-supervised learning strategy, where no fixed output $Y$ is attached to $X$, and the fitting function $F_a$ is evaluated with respect to a system of ``rewards'' and ``penalties'', so that the algorithm attempts to maximize the first and minimize the second. This type of machine learning is most often used when the algorithm needs to make a series of consecutive decisions to achieve a final goal (as for example when  attempting to win in a game such as chess or GO). Similarly to unsupervised learning, it has been shown (\cite{DRL}) that reinforcement learning works best when implemented in a modified, supervised learning setting.

Given the possibility of reducing both unsupervised and reinforcement learning to supervised learning schemes, we continue our analysis of supervised learning algorithms.

We now turn to consider another important question, suggested by the uncontrolled parameter structure of the supervised fitting function. Isn't it possible that, by working on a large enough data set, one can find arbitrary patterns in the data set that have no predictive power? The question can be addressed with the help of combinatorics, namely through Ramsey's theory. This makes possible to establish, in many cases, the minimal size for a set $\mathcal{S}$ in order to allow the identification of a given combinatorial pattern in a subset of  $\mathcal{S}$ \cite{ramsey}. By adapting Ramsey's theory to data analysis, Calude and Longo (\cite{CalLon17}) have shown that large enough data sets allow to establish any possible correlation among the data themselves.

This might suggest that wondering how much data is enough is only part of the story\footnote{On this question, note that there are ways to apply data science also to small data sets, if we accept strong limitations on the type of questions and we impose strong regularization restrictions on the type of solutions \cite{N}.}. Another, possibly more important one, if we want to avoid falling into the risk 
of making the result of blind methods perfectly insignificant, is wondering how much data is too much. 

There are several comments to be made on this matter. 

To begin with, we note that Ramsey's theory proves the existence of lower bounds on the size of sets of data that ensures the appearance of correlations. But these lower bounds are so large as to be of little significance for the size of data sets one usually handles. 

Even more importantly, Ramsey theory allows to establish the presence of patterns in subsets of the initial data set. In supervised learning, on the other hand, we require that every element of $X$  matches with an appropriate element of $Y$: this is essentially different from seeking correlations in a subset of the data set. In other words, Ramsey's theory would only show that it is possible to write $F(X_i)=Y_i$ for some specific subset of elements of $X$ and of $Y$: this would have no useful application in practice for supervised learning, where the totality of the available data must be properly matched.
 
Hence, as long as finding patterns within a data set $X$ is tied to supervised learning, there is no risk of uncontrolled and spurious correlations. Instead, any such correlation will be strongly dependent on its relevance in finding the most appropriate fitting function $F$. Moreover, we will see in Section 4 that, even when blind methods seem not to fall within the structural constraints of supervised learning, they can still be reinterpreted as such.

We should add that agnostic science enters the game not in opposition to traditional, theoretically-bound methods, but as an other mode of exploration of phenomena, and it should in no way discourage, or even inhibit the search for other methods based on previous understanding. 
Any form of understanding of the relevant phenomena is certainly welcome. Still, our point here is that there is no intrinsic methodological weakness in blind methods that is not, in a way or another, already implicit in those other methodologies with a theoretical bent. At their core they all depend on
some sort of inductive inference: the assumption that a predictive rule, or a functional interpolation of data, 
either justified by a structural account of phenomena, or by analytical continuation
of interpolating functions, will continue to hold true when confronted with new observations.

Supervised learning shows that we can succeed, despite the obvious (theoretical and/or
practical) risks, in using data to find pattern useful to solve specific problems with the available resources (though not necessarily patterns
universally associated with the relevant phenomena). The degree of success is manifest in disparate applications such as: face recognition \cite{FN}; automated translation algorithms \cite{GNTM}; playing (and beating humans) at difficult games such as GO \cite{DeepMindGO}; and even the noteworthy progress in self-driving cars \cite{SDC}. This makes agnostic science both useful and welcome.

It is, on the other hand, also because of the intrinsic and unavoidable risks of  agnostic science that it is important to understand why it works well so frequently, and what makes it 
successful. We should not be blind as to why blind methods succeed! Lack of understanding of phenomena does not necessarily require lack of understanding of agnostic science itself. Rather, it urgently asks
for such an understanding, in order to allow some sort of indirect (scientific, methodological, political or ethical) control possible.
This is the aim of an informed philosophy of data analysis, which shows, then, not only its intellectual
interest, but also its practical utility, and its necessity.

\section{Agnostic Science versus Lack of Control \label{sect.ASINS}}

Before continuing our search for such a (meta-)understanding, we observe that agnostic science is not equivalent to the use of data-driven algorithms on huge amount of data. As we will see in this section, we can single out  computationally efficient algorithms that can be applied to extremely large data sets.   And yet these very algorithms can be proven to converge, we fully understand their output, and we understand the structure of the data that makes them useful, so that they cannot considered blind algorithms, and their use is not an example of agnostic science. To better see this point, we will describe PageRank: the algorithm used by Google to weight webpages (\cite{BrinPage98},  \cite{PageBrinMotwani1998}).

Let $A$ be a web page with $n$ other pages  $T_{i}$ ($i= 1,2, \ldots , n$) pointing to it. We introduce a damping factor $d_{A}$ ($0 \leq d \leq 1$) that describes the probability that a random web-surfer landing on $A$ will leave the page.  If $d_{A}=0$, no surfer will leave the page  $A$;  if $d_{A}=1$,
every surfer will abandon the page.  One can chose $d_{A}$ arbitrarily, or on the basis of any possible a priori reason. Such choice does not affect the outcome of the algorithm in the limit of a sufficiently large number of iterations of the algorithm itself. The PageRank of $A$ is given by this formula:
$$
PR(A) = (1 - d_{A}) + d_{A}\left(\sum_{i=1}^{n} \frac{PR\left(T_i\right)}{C\left(T_i\right)}\right).
$$
where $PR\left(T_i\right)$ and $C\left(T_i\right)$ are respectively the PageRank of $T_i$ and the number of outgoing links starting at $T_i$.

This formula is very simple. But  it is recursive: in order to compute $PR(A)$, one needs to compute the PageRank  of all the pages pointing to $A$. In general, this makes impossible to directly compute it,
since, if $A$ points to some  $T_{i}$, then $PR\left(T_i\right)$ depends on $PR(A)$, in turn. However, this does not make the computation of $PR(A)$ impossible, since we can compute it by successive
approximations: {\it (1)} we begin by computing $PR(A)$ choosing any arbitrary value for $PR\left(T_i\right)$;  {\it (2)} the value  of $PR(A)$ computed in step {\it (1)}  is used
to provisionally compute $PR\left(T_i\right)$; {\it (3)} next  $PR(A)$ is recalculated on the basis of the values of  $PR\left(T_i\right)$ found in {\it (2)}; and so forth, for a sufficient number of times.

It is impossible to say a priori how many times the process is to be reiterated
in order to reach a stable value for any page of the Web. Moreover, the actual complexity and dimension of the Web makes impossible to follow the algorithm's computation at any of its stages, and for all the relevant pages. This is difficult even for a single page $A$, if the page is sufficiently connected within the Web. Since the Web is constantly changing, the PageRank of each page is not fixed and is to be computed again and again, so that the algorithm needs to be run continuously.  Thus it is obvious the impossibility of any local control on this process.

Still, it can be demonstrated that the algorithm converges to the principal eigenvector of the normalized link matrix of the Web. This makes the limit PageRank of any page, namely the value of the PageRank of the given page in this vector, a measure of the centrality of this page in the Web. 

Whether this actually measures the importance of the page is a totally different story.
What is relevant is that the algorithm has been designed to compute the principal eigenvector, under the assumption that the value obtained in this way is an index of the importance of the page.
Given any reasonable definition of importance, and under suitable conditions,  it has been recently proved (\cite{OlssonAl2016}, Theorem 2) that PageRank will asymptotically (in the size of the web) rank pages according to their importance.

This result confirms the essential point: the algorithm responds to a structural understanding of the Web, and to the assumption that the importance of any page is proportional to its centrally
in its normalized link matrix. Then, strictly speaking, there is nothing blind  in this approach, and using it is in no way an instance of agnostic science, though the Web is one
of the most obvious example of Big Data we might imagine. But, then, what makes blind methods blind, and agnostic science agnostic?

Agnostic science appears when, for the purpose of solving specify problems, one uses methods to search patterns which---unlike the case of PageRank---correspond to no previous understanding. This means we use methods and algorithms to find problem-dependent patterns in the hope that,  once discovered, they will provide an apparent solution to the given problem. If this is so, then agnostic science is not only a way to solve problems from data without structural understanding, but also a family of mathematically sophisticated techniques to learn from experience by observation of patterns.  Still, attention to invariant patterns is, ultimately, what Plato  (\textit{Theaetetus}, 155d) called
`astonishment [{\greektext{jaum'azein}}]', and considered to be ``the origin of philosophy [{\greektext{>arq`h filosof'ias}}]''. What happens with agnostic science is that we have too much data to be astonished by our experience as guided by the conceptual schemas we have at hand. So we use blind methods to  look for sources of astonishment deeply hidden within these data.

\section{Forcing and Deep Learning Neural Networks\label{sect.FOR}}

\subsection{Forcing Optimality}
We will now try to understand the features of an algorithm that make it suitable to identify appropriate patterns  within a specific problem.

The question has two facets. On one hand, it consists in asking what makes these algorithms successful.  On the other hand, it consists in wondering what makes them  so appropriate (for a specific
problem). The difficulty is that what appears to be a good answer to the first question  seems to contrast, at least at first glance, with the possibility of providing a satisfactory  answer to the second question.

Indeed, as to the first question, we would say, in our terminology, that the algorithms perform successfully because they act by forcing, i.e. by choosing interpolation methods and selecting functional spaces for the fitting functions in agreement to a criterion of intrinsic (mathematical) effectiveness, rather than conceiving  these methods in connection with the relevant phenomena. 

This answer seems to be in contrast with the possibility of providing a satisfactory answer to the second question, since it seems to negate from the start the possibility of understanding the appropriateness of methods in agnostic science. We think it is not. In this section we shall refine the notion of forcing by looking more carefully at the use of optimization in data science, and more particularly for a powerful class of algorithms, the so called Deep Learning Neural Networks.

Finally, in Section 5 we explore several ways to make the answer to the question of the successfulness of data science algorithms compatible with the existence of an answer to the appropriateness question. 

As we showed in \cite{nps1,nps5}, boosting algorithms are a clear example of forcing. They are designed to improve weak classifiers, generally just slightly better than random ones, and to transform them, by iteration, in strong classifiers. This is achieved by carefully focusing at each iteration on the data points that were misclassified at the previous iteration.
Boosting algorithms are particularly effective in improving the accuracy of classifiers based on sequences of binary decisions (so called ``classification trees''). Such classifiers are easy to build but on their own are relatively inaccurate. Boosting can, in some cases, reduce error rates for simple classification trees from  $45\%$ to about $5\%$ (\cite{Hastie2016}, Chapter 10).

Regularization algorithms offer a second example. If the data are too complicated and/or rough, these algorithms render them amenable to being treated by other algorithms, for example by reducing their dimension. Despite the variety of regularization algorithms, they can be all conceptually equated to the process of approximating a non-necessarily differentiable function by a function whose derivative absolute value is bounded from above everywhere on its domain.

The use of these algorithms reveals a double application of forcing: forcing on the original data to smooth them; and then forcing on the smooth data to treat them with a second set of algorithms. For example, after a regularization that forces the data to be smooth, some data science methods advocate the forcing of unjustified differential equations, in the search for a fitting function (\cite{fda}, chapter 19). These methods have been very effective in recognition of the authenticity of handwritten signatures, and they depend essentially on the condition that the data are accounted for by a smooth function.

Since in virtually all instances of forcing the mathematical structure of the methods is either directly, or indirectly reducible to an optimization technique, we claim that optimization is a form of forcing within the domain of agnostic science. 

In a sense, this is suggested by the historical origins of optimization methods (\cite{Panza1995}; \cite{Panza2003}). When Maupertuis, then President of the Berlin Academy of Sciences, first introduced the idea of least action, he claimed to have found the quantity that God wanted to minimize when creating the universe. Euler, at the time a member of the Berlin Academy of Sciences, could not openly criticize his President, but clearly adopted a different attitude, by maintaining that
action was nothing but what was expressed by the equations governing the system under consideration. 
In other terms, he suggested one should force the minimization (or maximization) of an expression like  $$ \int{F(x)dx}$$ on any physical system in order to find the function $F$ characteristic of it. \textit{Mutatis mutandis}, this is the basic idea that we associate
today with the Lagrangian of a system. Since then optimization became the preeminent methodology in solving empirical problems.  One could say that the idea of a Lagrangian has been generalized to the notion of fitting function,
whose optimization characterizes the dynamics of a given system.

Though this might be seen as a form of forcing, within a quite classical setting, one should note that, in this case, the only thing that is forced on the problem is the form of the relevant condition, and the request that a certain appropriate integral reaches a maximum or minimum. In this classical setting, however, 
the relevant variables are chosen on the basis of a preliminary understanding of the system itself, and the relevant function is
chosen so that its extremal values yield those solutions that have already been found in simpler cases. 

Things change radically when the fitting function is selected within a convenient functional space through an interpolation process designed to make the function fit the given data. In this case, both the space of functions and the specific fitting procedure (which makes the space of functions appropriate) are forced on the system. These conditions are often not enough
to select a unique fitting function or to find or ensure the existence of an absolute minimum, so that an additional choice may be required (forced)  to this purpose. 

There are many reasons why such an optimization process can be considered effective. One is that it matches the microarray principle: enough data, and a sufficiently flexible set of algorithms, will solve, in principle, any scientific problem. More concretely, optimization has shown to be both simple and relatively reliable: not necessarily to find the actual solution of
a problem, but rather to obtain, without exceeding time and resources constraints, outcomes that can be taken as
useful solutions to the problem. 
The outcomes of optimization processes can be  tested in simple cases and shown compatible with solutions that had been found with methods based  on a structural understanding of the relevant phenomenon\footnote{This is different from choosing the fitting function on the basis of solutions
previously obtained with the help of an appropriate understanding.}. In addition, the results of an optimization process may turn out to be suitable for practical purposes, even when it is not the best possible solution. 
An example is provided by algorithms for self-driving cars. In this case, the aim is not that of mimicking in the best possible way human reactions, but rather simply to have a car that can autonomously drive with a sufficient attention to the safety of the driver and of all other cars and pedestrians on the road. 

This last example makes it clear that we can conceive optimization as a motivation for finding algorithms without being constrained by the search for the best solution. Optimization becomes a conceptual framework for the development of blind methods. 

Blind methods are disconnected from any consideration of actual optimality; this sets them apart from methods in perturbation theory, where a solution to a more complex problem is derived from a (small) deformation of a solution of a simpler problem. On one hand, there is no doubt that looking at supervised learning as a case of interpolation leads naturally to a comparison with such a theory, and the sophistication of its most modern versions (including perturbation methods in quantum field theory\footnote{{For example, in \cite{NPS} a general classification problem from developmental biology is formulated as a path integral akin to those used in quantum mechanics. Such integrals are usually analyzed with the help of perturbative methods such as WKB approximation methods (see \cite{Schulman}, Chapter 18).}}) may  provide fundamental contributions to data science in this respect. On the other hand, blind methods place at the center the process of interpolation itself, rather than any correspondence between existing instances of solutions (i.e. simpler problems) and those to be determined (that we can equate to more complex problems).

Optimization as forcing also raises some important issues, beyond the obvious one which is typical of blind methods, namely the absence of an a priori justification.

One issue is that, in concrete data science applications such as pattern recognition or data mining, optimization techniques generally require fixing a large number of parameters, sometime millions of them, which not only make control of the algorithms hopeless, but also makes it difficult to understand
the way algorithms work. This large number of parameters often results in lack of robustness, since different initial choices of the parameters can lead to completely different solutions.

Another issue is evident when considering the default technique of most large scale optimization problems, the so called  point by point optimization. Essentially, this is a technique where the search for an optimal solution is done locally, by improving gradually any currently available candidate for the optimal choice of parameters.
This local search can be done, for example, by using the gradient descent method (\cite{DL}, section 4.3), which does not guarantee that we will reach  the desired minimum, or even a significant relative minimum. Since virtually all significant supervised machine learning methods can be
shown to be equivalent to point by point optimization \cite{nps5}, we will briefly describe the gradient descent method.

If $F(X)$ is a real-valued multi-variable function, its gradient $\nabla F$ is the vector that gives the slope  of its tangent oriented  towards the direction in which it increases most.
The gradient descent method exploits this fact to obtain a sequence  of values of $F$ which
converges to a minimum. Indeed, if we take $K_n$ small enough and set
$$
x_{n+1} = x_{n} - K_{n} \nabla F\left(x_{n}\right)~~(x=0,1, \ldots)
$$
then
$$
F\left(x_{0}\right) \geq F\left(x_{1}\right) \geq F\left(x_{2}\right), \ldots
$$
and one can hope that this sequence of values converges towards the desired minimum. However, this is only a hope, since nothing in the method can warrant that the minimum it detects is
significant.

\subsection{Deep Learning Neural Networks}

Let us now further illustrate the idea of optimization as forcing, by considering the paradigmatic example of Deep Learning Neural Networks (we follow here \cite{Hastie2016}, Section 11.3, and \cite{DL}) as it applies to the simple case of classification problems.

The basic idea is the same anticipated above for the search of a fitting function $F$ by supervised learning. One starts with a Training Set $\left(X,Y\right)$, where $X$ is a collection of $M$ arrays of variables:
$$
X = \left(X_{1}, \ldots , X_{M}\right) \qquad X_{i} = \left(X^{[1]}_{i}, \ldots X^{[N]}_{i}\right)
$$
and $Y$ is a corresponding collection of $M$ variables:
$$
Y = \left(Y_{1}, \ldots , Y_{M}\right).
$$
What is specific to deep learning neural networks is the set of specific steps used to recursively build $F$.

\begin{enumerate}
\item We build $K$ linear functions 
$$
Q^{[k]}(X_i)= A_{0,k} + \sum_{n=1}^{N}A_{n,k}X^{[n]}_{i} \qquad (k = 1, \ldots , K),
$$
where $A_{n,k}$ are $K(N+1)$ parameters chosen on some a priori criterion, possibly even randomly, and
$K$ is a positive integer chosen on the basis of the particular application of the algorithm.

\item One then selects an appropriate non-linear function $G$ (we will say more about how this function is chosen) to obtain $K$ new arrays of variables
$$
H^{[k]}(X_i)
= G \left(Q^{[k]}(X_i)\right) \qquad (k = 1, \ldots , K).
$$

\item One chooses (as in step 1) a new set of $T(K+1)$ parameters $B_{k,t}$ in order to obtain $T$ linear combinations of the variables $H^{[k]}(X_i)$
$$
Z^{[t]}(X_i) = B_{0,t} + \sum_{k=1}^{K}B_{k,t} H^{[k]}(X_i)\qquad (t = 1, \ldots , T),
$$
where $T$ is a positive integer appropriately chosen in accordance with the particular application of the algorithm.
\end{enumerate}
 If we stop after a single application of steps 1-3, the neural network is said to have only one layer (and is, then, ``shallow'' or not deep). We can set $T=1$ and the process ends by imposing that all the values of the parameters are suitably modified (in a way to be described shortly) to ensure that
$$
Z^{[1]}(X_i) \approx Y_{i}\qquad (i = 1, \ldots , M),  
$$
For any given new input $\tilde X$, we can then define our fitting function $F$ as $F(\tilde X)=Z^{[1]}(\tilde X)$\footnote{For classification problems, one often imposes $P(Z^{[1]}(X_i)) \approx Y_{i}, (i = 1, \ldots , M)$, where $P$ is a final, suitably chosen, output function (see \cite{Hastie2016}, Section 11.3). For any new input $\tilde X$, the fitting function is, then, $F(\tilde X)=P(Z^{[1]}(\tilde X))$.}. In deep networks, steps 1-3 are iterated several
times, starting every new iteration from the $M$ arrays $Z^{[t]}(X_i)$ constructed by the previous iteration. This iterative procedure creates several ``layers'', by choosing different parameters $A$ and $B$ (possibly of different size as well) at each iteration, with the obvious limitation that the dimension of the output of the last layer $L$ has to match the dimension of the elements of $Y$. If we denote by $Z^{[1]}_L(X_i)$ the output of the last layer $L$, we impose, similarly to the case of one layer, that $Z^{[1]}_L(X_i)\approx Y_{i}$.

In other words, the construction of a deep learning neural network involves the repeated transformation of an input X by the recursive application of a linear transformation (step 1) followed by a nonlinear transformation (step 2) and then another linear transformation (step 3).

The algorithm is designed to allow learning also in absence of $Y$ by using $X$ itself, possibly appropriately regularized, in place of $Y$ (auto-encoding). When an
independent $Y$ is used, the learning is called `supervised' and provides an instance of the setting described in Section~\ref{sect.ASINS}. In absence of it, the learning is,
instead, called `unsupervised' (\cite{DL}, chapter 14), and its purpose is to find significant patterns and correlations within the set $X$ itself. The possibility of using an algorithm designed for supervised learning also for unsupervised learning is an important shift of perspective. It allows to constrain the exploration of patterns within $X$, for the sole purpose of regularizing the data themselves. Whichever correlations and patterns are found, they will be instrumental to this specific  aim, rather than to the ambiguous task of finding causal relationships within $X$.

Two things remain to be explained. The first concerns the non-linear function $G$, called `activation function' (because of the origin of the algorithm as a  model for neural dynamic). Such function can take different forms. Two classical examples are the sigmoid function
$$
G(u) = \frac{1}{1+e^{-u}}
$$
and the ReLU (Rectified Linear Unit) function
$$
G(u)=max(0, u).
$$
This second function is composed of two linear branches and therefore is, mathematically speaking, much simpler than the sigmoid function. While the ReLU is also not linear, it has uniform slope on a wide portion of its domain, and this seems to explain its significantly better performance as activation function for deep networks. The use of an appropriate activation function allows the method to approximate any function that is continuous on the compact sets in $\mathbb{R}^n$. This result is known as the Universal Approximation Theorem for Neural Networks (\cite{hornik})

The second thing to be explained concerns the computation of the parameters according to the condition:
$$
Z^{[1]}_L(X_i) \approx Y_{i}\qquad (i = 1, \ldots , M).
$$
This is typically achieved through the gradient descent method by minimizing a fitness function such as:
$$
\sum_{i=1}^{M}\left[Y_{i} - Z^{[1]}_L(X_i)\right]^2.
$$

The gradient is computed by an appropriate fast algorithm adapted to neural networks known as
backpropagation (\cite{Hastie2016}, Section 11.4). As we already noted in \cite{nps3}, the effectiveness of neural networks (both deep and shallow) seems to depend more on the specific structure of the backpropagation algorithm, than on the universal approximation properties of neural networks.  Note also that the minimization of the fitness function is equivalent to a regularization of the final fitting function, if we stop the iterative application of the backpropagation algorithm when the value of the fitness function does not significantly decreases any more (\cite{DL}, Section 7.8).
 
When dealing with deep networks, one can go as far as considering hundreds of layers, though it is not generally true that increasing the number of layers always improves the minimum of the corresponding fitness function.
Nevertheless, in many cases, taking more layers often allows up to a tenfold reduction of errors. For example, it has been showed that  in a database of images of handwritten digits, classification errors go from a rate  
of 1.6\% for a 2 layers network (\cite{Hastie2016}, section 11.7) to a rate of  0.23\% with a network of about 10 layers (\cite{10layers}).

This short description of the way in which Deep Learning Neural Networks work should be enough to clarify why we have taken them as an example of optimization by forcing. Above all, both the dependence of the effectiveness of neural networks on the structure of the backpropagation algorithm, and  their interpretation as regularization are clear indications that the way neural networks are applied is an instance of forcing.

More broadly, the opacity of the recursive process that creates the layers of the network is matched by computationally driven considerations that establish the specific type of gradient descent method to be used, and by a criterion to stop the whole iterative process that is simply based on the inability to find better solutions.
But this same description should also be enough to make clear the second question mentioned in the beginning of the present section: how can methods like deep learning neural networks be appropriate for solving specific problems, when the methods themselves do not reflect in any way the particular features of the problems? We explore this question in the next section.

\section{ On the Appropriateness of Blind Methods  \label{sect.Con}}
\subsection{Understanding Methods rather than Phenomena }
A simple way to elude the question of the appropriateness of blind methods is by negating its  premise: one can argue that, in fact, blind methods are in no way appropriate; that their success is nothing but appearance and that the faith in their success is  actually dangerous, since such faith provides an incentive to the practice of accepting illusory forecasts and solutions.

The problem with this argument is that it  ultimately depends on arguing that blind methods do not succeed since they do not conform with the pattern of classical science.
But an objective look at the results obtained in data science should be enough to convince ourselves that this cannot be a good strategy. Of course, to come back to the example of DNA microarrays, grounding cancer therapy only on microarrays is as inappropriate as it is dangerous, since, in a domain like that, looking for causes is as crucial as it is necessary. But we cannot deny the fact that microarrays can be used also as an evidential bases in a search for causes. And also we cannot deny that, in many successful applications of blind methods---like in handwriting recognition---the search of causes is much less crucial. 

So we need another justification of the effectiveness of blind methods, which, far from negating the appropriateness question, takes it seriously and challenges the assumption that classical science is the only appropriate pattern for good science. Such an approach cannot depend, of course, on the assumption that blind methods succeed because they perform appropriate optimization.  This assumption merely displaces the problem, since optimization is only used by these methods as a device to find possible solutions.

A more promising response to the question of the appropriateness of blind methods might be that they succeed for the same reason as classical induction does: blinds methods are indeed interpolation methods on input/output pairs,
followed by analytical continuation, which is how induction works. Of course, one could argue that induction itself is not logically sound. But could one really reject it as an appropriate method in science
because of this? Is there another way to be empiricist, other than trusting induction?  
And can one really defend classical science without accepting some form of empiricism, as refined as it might be? We think all these questions should be answered in the negative, and therefore we believe the objection itself to be immaterial. 

There are, however, two other important objections to this response.

The first is that it applies only to supervised methods, that is, methods based on the consideration of a training set on which interpolation is performed. It does not apply, at least not immediately, to unsupervised ones, where no sort of induction is present.  However, this objection is superseded by noting that it is possible to reduce unsupervised methods to supervised ones through the auto-encoding regularization processes described above.

The second objection is more relevant. It consists in recognizing that, when forcing is at work, interpolation is restricted to a  space of functions which is not selected by considering the specific nature of the relevant phenomenon, and that cannot be justified by any sort of induction. The choice of the functional space, rather, corresponds to a regularization of the data and it often modifies those data in a way that does not reflect, mathematically, the phenomenon itself. 

This objection is not strong enough to force us to completely dismiss the induction response. But it makes it clear that advocating the powerfulness of induction cannot be enough to explain
the success of agnostic science. This general response is, at least, to be complemented by a more specific and stronger approach.

In the remainder of this section, we would like to offer the beginning of a new perspective, which is consistent with our interpretation of the structure of blind methods.

The basic idea is to stop looking at the appropriateness question as a question concerning some kind of correspondence between phenomena and  specific solutions found by blind methods. The very use of forcing makes this perspective illusory.  We should instead look at the question from a more abstract, and structural perspective. Our conjecture, already advanced in \cite{nps4,nps5}, is that we can find
a significant correspondence between the structure of the algorithms used to solve problems, and the way in which phenomena of 
interest in data science are selected and conceived. We submit, indeed, that the (general) structure of blind methods, together with the formal features of the Microarray Paradigm, exert strong restrictions on the class of data sets that agnostic science deals with.

\subsection{ Brandt's Principle and the Dynamics of Blind Methods}

To justify the existence of these restrictions, we start by recalling a result of \cite{nps5}, that all blind methods share a common structure that conforms to the following prescription: 
\begin{quote}
{\it An algorithm that approaches a steady state in its output has found a solution to a problem, or needs to be replaced.}
\end{quote}
\noindent
In \cite{nps5} we called this prescription `Brandt's Principle', to reflect the fact that it was first expounded by  Achi Brandt for the restricted class of multiscale algorithms (\cite{Brandt2002}).

As simple as Brandt's Principle appears at first glance, in \cite{nps5} we showed that this principle allows a comprehensive and coherent reflection the structure of blind methods in agnostic science. First of all,  Brandt's Principle is implicit in forcing, since an integral idea in forcing is that if an algorithm 
does not work, another one is to be chosen. 
But, more specifically, the key to the power of this principle is that the steady state output of each algorithm, when it is reached, is chosen as input of the next algorithm, if a suitable solution to the initial problem has not yet been found.

Notably, deep learning architecture matches with Brandt's Principle, since the iteration of the gradient descent algorithm is generally stopped when the improvement of parameters reaches a steady state and, then, either the function that has been obtained is accepted and used for forecasting or problem-solving, or the algorithm is replaced by a new one, or at least re-applied starting from a new assignation of values to the initial parameters. More than that, all local optimization methods satisfy this principle. And since most algorithms in agnostic data science can be rewritten as local optimization  methods, we can say that virtually all algorithms in agnostic data science do. 

In \cite{nps5} we argued that thinking about algorithms in terms of Brandt's principle often sheds light on those characteristics of a specific method that are essential to its success. For example, the success of deep learning algorithms, as we have seen in the previous section, relies in a fundamental way on two advances: (1)  the use of the ReLU activation function that, thanks to its constant slope for nonzero arguments, allows the fast exploration of the parameter space with gradient descent; and (2)
a well defined regularization obtained by stopping the gradient descent algorithm when error rates do not improve significantly anymore. Both these advances took a significant time to be identified as fundamental to the success of deep learning algorithms, perhaps exactly for their deceiving simplicity, and yet both of them are naturally derived from Brandt's principle.

There is, however, a possible objection against ascribing a decisive importance to this principle (one that is in the same vein as that we discussed in section \ref{sect.MP}, considering an argument from \cite{CalLon17}). This objection relies on the observation that, in practical applications, agnostic science works with floating-point
computations, which require a finite set of floating-point numbers. The point, then, is that any iterative algorithm on a finite set of inputs reaches a limit cycle in finite time, in which case also steady states 
satisfying Brandt's principle become trivial and uninformative about the nature of the subjacent phenomenon. 

However, blind methods that satisfy Brandt's principle, such as boosting algorithms and neural networks, will usually converge to steady states after just a few thousands iterations. Limit cycles in an algorithm's output due to the limitations of floating-point arithmetic will instead appear after a very large number of iterations, comparable to the size of the set of floating-point numbers. Any practical application of Brandt's principle needs to take this into  consideration, by imposing, for example, that the number of iterations necessary to reach a steady state is at most linear in the size of the training set.

Regardless of this practical limitation in recognizing steady states, the real significance of Brandt's principle for supervised learning algorithms is that it shifts the attention from the optimization of the fitting function $F$ to the study of the dynamics of the algorithms' output. In this perspective, applying Brandt's Principle depends
on building sequences of steady-state fitting functions $\{F_j\}$  that are stable in their performance under the deformations induced by the algorithms
that we choose to use during the implementation of the principle itself. This generates 
a space of fitting functions, that are mapped into each other by the different algorithms. 

A clear example of this process is provided by boosting, where the entire family of fitting functions (that is recursively found) is robust  in its performance, once the classification error on the training set stabilizes. Moreover, fitting functions found by the algorithm at each recursive step are built by combining all those that were found earlier (by weighing and adding them suitably). Indeed, boosting is an instance of ``ensemble methods'', where distinct fitting functions are combined to find a single final fitting function. It  can be argued that a lot of the recent progress in data science has been exactly due to the recognition of the central role of ensemble methods, such as boosting, in greatly 
reducing error rates when solving problems (\cite{Hastie2016}, Chapter 16).

\subsection{ Ensemble Methods and Interpolation}

Note that for ensemble methods to be trustworthy, eventually they must stabilize to a fitting function that does not significantly change with the addition of more sample points. Therefore, if we take the limit of an infinite number of sample points, generically distributed across the domain where the relevant problem is well defined, the fitting function is forced to be unique.

To understand the significance of this instance of forcing, we first rephrase the basic slogan of the Microarray Paradigm in more precise terms as a `Quantitative Microarray Paradigm':
\begin{quote}
{\it Given enough sample data points, and for a large and diverse enough set of variables $X$, the value of any  other variable $Y$ relevant for the solution of a given problem can be generally calculated from the value of $X$ (via the fitting function $F(X)=Y$).} 
\end{quote}
\noindent
Once this assumption is admitted, the unicity of $F$ in the limit entails a form of analyticity on the fitting function which we call `asymptotic sample-analyticity':
\begin{quote}
{\it Let $N$ be the dimension of $X$; then, for $N$ sufficiently large, $F(X)=Y$ is uniquely determined, on an appropriate domain, by a generic infinite set of sample points.}
\end{quote}
In application, we will always have finite data, and we will not be able to choose $F$ uniquely to solve a problem. But suppose that the same solution is given by the entire class of asymptotically sample-analytic functions that are compatible with the available data. Do we trust that such solution is reflective of a relevant actual property of the phenomenon at issue\footnote{Note that sample-analyticity can be forced on any discrete variable in the problem after imposing continuity on such variables.}? The question shifts the attention from the nature of data sets to the nature of the space of functions defined on them, and on their assemblage by appropriate algorithms.

Our suggestion is that blind methods succeed when (and because) they select, in agreement with Brandt's principle, appropriate classes of
asymptotically sample-analytic functions, apt to robustly provide uniquely determinate solutions for the data problems at issue. 

Despite this shift from data sets to functions on data sets, the properties of such functions enforce some general conditions on the data as well. First of all, the Quantitative Microarray Paradigm essentially requires variables in the data set to be strongly interdependent in the limit of large data sets. Second, we claimed at the end of Section 5.2 that Brandt's Principle identifies a space (ensemble) of fitting functions that are stable in their ability to solve a given problem. Such stability is possible only if the functional relations that can be defined on the variables are themselves robust, so that they persist across the large data sets that are required by the microarray paradigm. It is not important for these interdependence and robustness to be apparent, that is, we do not need to be able to identify the specific dependence of variables from each other.
Nor it is important for robustness to warrant a long-term conservation of specific relations among the variables. What is relevant is that the interdependence is strong and persistent enough to allow iterative algorithms conforming to Brandt's Principle to subsequently correct their outputs by building more and more convenient fitting functions.

The requirements of 
interdependence of variables and robustness of functional relations among them can now be used to discriminate data sets most suitable for the application of blind methods.
For example, such requirements are believed to be satisfied by developmental biological systems (see \cite{minelliForm}). Moreover,  in \cite{nps5} we gave evidence that social and economical systems satisfy a generalization of the `principle of Developmental Inertia', an organizing principle for developmental biology (first proposed in \cite{minelli2011}). It is therefore likely that the same interdependence and robustness satisfied by biological systems hold for most social and economical systems as well.

We conclude that the Microarray Paradigm and Brandt's principle enforce specific requirements on data sets and that these requirements are likely to be satisfied by data arising from biological, social and economical systems. Blind methods would then be most appropriate when applied to such systems.

\section{Conclusion}

In this paper we reviewed and extended a perspective on the methodological structure of data science which we have been building in a series of papers \cite{nps1, nps3,nps4,nps5}. The basic assumption of our approach is that data science is a coherent approach to empirical problems that in its most general form does not build understanding about phenomena. Because of this characteristic, we labelled this approach to empirical phenomena `agnostic science', and called the methods that make up agnostic science `blind methods'.

The basic attitude underlying agnostic science is the belief that if enough and sufficiently diverse data are collected regarding a certain phenomenon, it is possible to answer all relevant questions about it. In Section 2 we referred to this belief as the microarray paradigm and we explored the specific ways it is used in the practice of machine learning.

We noted in Section 3 that not all computational methods dealing with large data sets are properly within the domain of agnostic science, and we gave the example of PageRank, an algorithm used to weight webpages. The convergence of this algorithm and the significance of its output are readily intelligible and therefore we argued that PageRank is not a blind method.

In Section 4.1 we explored how the microarray paradigm calls for a new type of mathematization in agnostic science, where mathematical methods are forced on the problem, i.e., they are applied to a specific problem only on the basis of their ability to reorganize the data for further analysis by general purpose techniques that are selected only on the basis of the richness of their mathematical structure, rather than by any particular relevance for the problem at hand. We then showed that optimization methods are used in data science as a form of forcing. This is particularly significant since virtually all methods of data science can be rephrased as a type of optimization method. In particular, in Section 4.2 we argued that deep learning neural networks are best understood within the context of forcing optimality. 

In Section 5 we moved to the broader question of the appropriateness of blind methods in solving problems. In Section 5.1 we argued that this question should not be interpreted as a search for a correspondence between phenomena and  specific solutions found by blind methods. Rather, it is the internal structure of blind methods that should be understood, and its implications on the structure of the data sets that are most appropriate for such methods.

To this extent, we reviewed in Section 5.2 a simple prescription on algorithms, Brandt's principle, which asserts that an algorithm that approaches a steady state in its output has found a solution to a problem, or needs to be replaced. One of our main claims in \cite{nps5} was that Brandt's principle is ideally suited to the understanding of the dynamical structure of blind methods. For example, in Section 5.2 we used Brandt's principle to understand two of the significant innovations of deep learning neural networks: the use of the ReLU activation function in the network; and an efficient criterion for early stopping of the algorithm. Ensemble methods, where distinct fitting functions are combined to find a single final fitting function, can also be interpreted within the context of Brandt's principle. 

In Section 5.3 we showed that Brandt's principle and the microarray paradigm force a specific type of analytical structure, which we call `sample-analyticity', on the final fitting function found by ensemble methods. And we argued that sample-analyticity forces a shift from data sets to functions on data sets. In turn, the properties of such functions enforce two general conditions on the data sets: a strong interconnectedness of the variables of the data set; and the robustness of the functional relations of such variables. 

We finally speculated that blind methods are most appropriate for the solution of problems in biological, social and economical systems, since data sets arising from these systems are likely to satisfy the two conditions above.

\section*{Acknowledgments}

We thank Maxime Darrin for reading and commenting a preliminary version of this paper and the anonymous referees for many detailed and useful remarks.

\end{document}